# Klein Tunneling and Supercollimation of Pseudospin-1 Electromagnetic Waves

A. Fang[1,2], Z. Q. Zhang[1,2], Steven G. Louie[2,3,4] and C. T. Chan[1,2,*]

[1]Department of Physics, The Hong Kong University of Science and Technology, Clear Water Bay, Hong Kong, China

[2]Institute for Advanced Study, The Hong Kong University of Science and Technology, Clear Water Bay, Hong Kong, China

[3]Department of Physics, University of California at Berkeley, Berkeley, CA 94720, USA

[4]Materials Sciences Division, Lawrence Berkeley National Laboratory, Berkeley, CA 94720, USA

*Email: phchan@ust.hk

Pseudospin plays a central role in many novel physical properties of graphene and other artificial systems which have pseudospins of ½. Here we show that in certain photonic crystals (PCs) exhibiting conical dispersions at k = 0, the eigenmodes near the "Dirac-like point" can be described by an effective spin-orbit Hamiltonian with a higher dimension value $S$=1, treating the wave propagation in positive index (upper cone), negative index (lower cone) and zero index (flat band) media within a unified framework. The 3-component spinor gives rise to boundary conditions distinct from those of pseudospin-½, leading to new wave transport behaviors as manifested in super Klein tunneling and supercollimation. For example, collimation can be realized more easily with pseudospin-1 than pseudospin-½. The effective medium description of the PCs allows us to further understand the physics of pseudospin-1 electromagnetic (EM) waves from the perspective of complementary materials. The special wave scattering properties of pseudospin-1 EM waves, in conjunction with the discovery that the effective photonic potential can be varied by a simple change of length scale, offer new ways to control photon transport. As a useful platform to study pseudospin-1 physics, dielectric PCs are much easier to fabricate and characterize than ultracold atom systems proposed previously. The system also provides a platform to realize the concept of "complementary medium" using dielectric materials and has the unique advantage of low loss.

# I. INTRODUCTION

In the past decade, graphene has become a fruitful platform to study various novel physical phenomena in condensed matter physics and material science,[1-15] such as Klein tunneling,[7,15] Zitterbewegung,[8,9,15] integer quantum Hall effect,[4-6] weak antilocalization,[10,11] and supercollimation of electron beams.[12-14] Most of these properties can be attributed to its unique conical band structure at the Brillouin zone boundary ($K$ and $K'$ points), with two cones meeting at one point at the Fermi level. The low energy excitations can be described by a massless Dirac equation with its wave function represented by a two-component spinor.[15-17] Each component of the spinor corresponds to the amplitude of the wave function on one trigonal sublattice of graphene. As such, graphene is frequently considered as a "pseudospin-½" system. This pseudospin-½ spinor is aligned either parallel or antiparallel to the wavevector **k**. A natural question to ask is whether one can construct systems with a higher pseudospin value and whether there exist any interesting physics that are distinct from those observed in pseudospin-½ systems. Previous work[18-22] suggested that one can indeed achieve pseudospin $S = 1$ in ultracold atom systems, which can lead to striking transport properties, such as super Klein tunneling, i.e., perfect transmission for all incident angles. However, as such systems demand extremely low temperature and are technically difficult to realize experimentally, the predicted novel behaviors would be very challenging to observe experimentally. In this work we demonstrate that the photon transport in certain PCs is also governed by a pseudospin-1 Hamiltonian, which offers the opportunity to study the related physics in material platforms that are much easier to realize experimentally. And we call such PCs "photonic pseudospin-1 materials".

For photonic systems, conical dispersion can also exist at **k** = 0. It was demonstrated[23-27] that in some 2D dielectric photonic crystals a Dirac-like cone can occur at the center of Brillouin zone (**k**=0). The conical dispersions are different from the Dirac cones in graphene, as they are always accompanied by an additional flat band crossing the Dirac-like point (see Fig. 1). The triply degenerate states at the Dirac-like point are formed by the accidental degeneracy of monopole and dipole excitations.[23-26] The PCs can be described by an effective medium with simultaneously zero permittivity and permeability[23-26] at the Dirac-like point frequency, and the effective index is positive/negative at frequencies above/below the Dirac-like point frequency. Some authors[28] describe the property of such double-zero materials using $2\times 2$ matrices and ignore the existence of the flat band. However, this omission is actually not acceptable because a Dirac cone corresponding to pseudospin-½ gives rise to a Berry phase of π, whereas the Berry phase of a Dirac-like cone in the present case is zero.[23,25]

In this work we will demonstrate that the conical dispersion of PCs near **k** = 0, together with the flat band, can be mapped into an effective spin-orbit Hamiltonian with a pseudospin of 1. This Hamiltonian describes the wave transport properties in both positive and negative refractive index regimes which correspond to the upper and lower conical bands, respectively, within a unified framework. The three components of pseudospin-1 EM waves are related to three independent electromagnetic modes for each **k**-point near the Dirac-like point: one longitudinal mode (its existence made possible by a zero effective index), one transverse mode with the positive dispersion and another transverse mode with the negative dispersion. These three pseudospin states represent a new basis for pseudospin-1 EM waves, and are different from and additional to the intrinsic spin states which are used to describe the helicity of a circularly polarized photon in a homogeneous medium. The pseudospin-1 state can only exist in a medium with an underlying subwavelength lattice structure.

Before we discuss the interesting wave transport behaviors of pseudospin-1 EM waves in pseudospin-1 materials, we need to address the problem of controlling wave transport in such systems. For graphene, transport can be conveniently controlled by applying a gate voltage. What then is the corresponding mechanism that can effectively shift the "photonic potential"? We will show below that a simple change of length scale of the PC can shift the Dirac-like cones up or down in frequency with its group velocity unchanged, hence mimicking a gate voltage in graphene and other charged Dirac fermion systems. As an analogy with electron potential, we can map the shift of the Dirac-like frequency to a change of "photonic potential" within the effective Hamiltonian description near the Dirac-like frequency. With the availability of effective photonic potential to control wave transport, we studied the transport properties of pseudospin-1 EM waves in 1D photonic potential systems. Full-wave numerical simulations show that we can indeed achieve the predicted super Klein tunneling through a square photonic potential barrier, realized by a sandwich structure composed of two PCs with different length scales. Moreover, we find very robust supercollimation of pseudospin-1 EM waves which can be achieved in any Kronig-Penney type of potential consisting of layers of equal thickness. In contrast, for electrons, supercollimation is achievable only in some specific superlattice potentials.[13]

## II. RESULTS

### A. Spin-orbit Hamiltonian near the Dirac-like cone from Maxwell's equations

For a 2D problem, we consider a transverse electric (TE) polarized EM wave propagating in the $xy$ plane. The TE solution has electric field $\mathbf{E} = (0, 0, E_z)$ and magnetic field $\mathbf{H} = (H_x, H_y, 0)$, and the material constitutive tensors are permittivity $\vec{\varepsilon} = \mathrm{diag}(\varepsilon_{xx}, \varepsilon_{yy}, \varepsilon_{zz})$ and permeability $\vec{\mu} = \mathrm{diag}(\mu_{xx}, \mu_{yy}, \mu_{zz})$. Note here that only $\varepsilon_{zz}$, $\mu_{xx}$ and $\mu_{yy}$ enter the problem. Setting $\varepsilon_{zz} = \varepsilon$ and $\mu_{xx} = \mu_{yy} = \mu$, where $\varepsilon$ and $\mu$ are functions of frequency $\omega$, we can write the Maxwell's equations as a matrix equation (see Appendix A),

$$\begin{pmatrix} 0 & -i\frac{\partial}{\partial x} - \frac{\partial}{\partial y} & 0 \\ -i\frac{\partial}{\partial x} + \frac{\partial}{\partial y} & 0 & -i\frac{\partial}{\partial x} - \frac{\partial}{\partial y} \\ 0 & -i\frac{\partial}{\partial x} + \frac{\partial}{\partial y} & 0 \end{pmatrix} \tilde{\psi} = \omega \begin{pmatrix} \mu & 0 & 0 \\ 0 & 2\varepsilon & 0 \\ 0 & 0 & \mu \end{pmatrix} \tilde{\psi}, \qquad (1)$$

where $\tilde{\psi}$ is $\tilde{\psi}_T = (-iH_x - H_y, E_z, iH_x - H_y)^T$ for transverse modes and $\tilde{\psi}_L = (H_x - iH_y, E_z, -H_x - iH_y)^T$ with $E_z$ being spatially independent for longitudinal modes, respectively. For PCs with a conical dispersion at $\mathbf{k} = 0$, the effective permittivity $\varepsilon$ and permeability $\mu$ vanish simultaneously at the Dirac-like point frequency, $\omega_D$, i.e., $\varepsilon(\omega_D) = \mu(\omega_D) = 0$.[24,25] In the neighborhood of $\omega_D$, $\omega\varepsilon$ and $\omega\mu$ can be approximated by

$\omega\varepsilon \cong \omega_D \left.\frac{d\varepsilon}{d\omega}\right|_{\omega=\omega_D} (\omega - \omega_D) \equiv (\omega - \omega_D)\tilde{\varepsilon}$ and $\omega\mu \cong \omega_D \left.\frac{d\mu}{d\omega}\right|_{\omega=\omega_D} (\omega - \omega_D) \equiv (\omega - \omega_D)\tilde{\mu}$. Here both $\tilde{\varepsilon}$ and $\tilde{\mu}$ are positive definite.[24, 25] With the above linear expansion, we can rewrite Eq. (1) in $k$-space as

$$\frac{1}{\sqrt{2\tilde{\varepsilon}\tilde{\mu}}} \begin{pmatrix} 0 & k_x - ik_y & 0 \\ k_x + ik_y & 0 & k_x - ik_y \\ 0 & k_x + ik_y & 0 \end{pmatrix} \psi^{(\mathbf{k})} = \delta\omega\psi^{(\mathbf{k})}, \qquad (2)$$

where $\delta\omega = \omega - \omega_D$ and $\psi^{(\mathbf{k})} = U^{-1}\tilde{\psi}^{(\mathbf{k})}$, with $\tilde{\psi}^{(\mathbf{k})} = \frac{1}{2\pi}\int \tilde{\psi} e^{-i\mathbf{k}\cdot\mathbf{r}} d\mathbf{r}$ and

$U = \text{diag}\left(2\sqrt{\tilde{\varepsilon}/\tilde{\mu}},\ \sqrt{2},\ 2\sqrt{\tilde{\varepsilon}/\tilde{\mu}}\right)$ (see Appendix B). It should be pointed out that at frequencies below or above $\omega_D$, the system is effectively a negative ($\varepsilon,\ \mu < 0$) or positive ($\varepsilon,\ \mu > 0$) refractive index medium, respectively. The change of signs in $\varepsilon$ and $\mu$ in Eq. (1) as the frequency moves across $\omega_D$ is now directly reflected in the sign change of $\delta\omega$ in Eq. (2). Thus, equation (2) treats the positive and negative index behaviors of the system in a unified framework. If we use the following matrix representation of spin-1 operator, i.e., $\mathbf{S} = S_x \hat{x} + S_y \hat{y}$ with

$$S_x = \frac{1}{\sqrt{2}}\begin{pmatrix} 0 & 1 & 0 \\ 1 & 0 & 1 \\ 0 & 1 & 0 \end{pmatrix},\quad S_y = \frac{1}{\sqrt{2}}\begin{pmatrix} 0 & -i & 0 \\ i & 0 & -i \\ 0 & i & 0 \end{pmatrix}, \tag{3}$$

equation (2) can be written as $H\psi^{(\mathbf{k})} = \delta\omega\psi^{(\mathbf{k})}$ with

$$H = v_g \mathbf{S}\cdot\mathbf{k}, \tag{4}$$

where $v_g = 1/\sqrt{\tilde{\varepsilon}\tilde{\mu}}$, is the group velocity of the Dirac-like cone (see Appendix C). Thus, equation (2) or (4) represents a spin-orbit interaction with pseudospin-1. Equation (4) is also mathematically equivalent to the Hamiltonian of a magnetic dipole moment in a magnetic field, with $-v_g \mathbf{S}$ and $\mathbf{k}$ playing the respective roles of the magnetic dipole moment and magnetic field. Since $\mathbf{k}$ is in the $xy$ plane, three normalized eigenvectors have the following forms:

$$\bar{\psi}_s^{(\mathbf{k})} = \frac{1}{2}\begin{pmatrix} se^{-i\theta_k} \\ \sqrt{2} \\ se^{i\theta_k} \end{pmatrix} (s = \pm 1) \text{ and } \bar{\psi}_s^{(\mathbf{k})} = \frac{1}{\sqrt{2}}\begin{pmatrix} e^{-i\theta_k} \\ 0 \\ -e^{i\theta_k} \end{pmatrix} (s = 0), \tag{5}$$

where $\theta_{\mathbf{k}}$ is the angle of the wavevector $\mathbf{k}$ with respect to the $x$-axis. The corresponding eigenvalues are

$\delta\omega = sv_g|\mathbf{k}|$, where $s = 0, \pm 1$ denote the upper conical band ($s = +1$), the flat band ($s = 0$) and the lower conical band ($s = -1$), respectively. They also describe the pseudospin states which are parallel ($s = +1$), perpendicular ($s = 0$) and anti-parallel ($s = -1$) to the wavevector $\mathbf{k}$. Thus, the upper and lower cones have opposite chiralities. It is easy to show that the Berry phase, $\gamma = i\oint \langle \bar{\psi}_s^{(\mathbf{k})}|\nabla_{\mathbf{k}}|\bar{\psi}_s^{(\mathbf{k})}\rangle \cdot d\mathbf{k} = 0$ for all three bands, consistent with the previous results.[23,25] It should be mentioned that the two normalized eigenvectors $\bar{\psi}_s^{(\mathbf{k})}$ ($s = \pm 1$) in Eq. (5) give rise to two transverse modes with one in a positive medium and the other in a negative medium as can be seen below. In the TE modes considered here, the electric field $E_z^{(\mathbf{k})}$ of the wavevector $\mathbf{k}$ in the normalized eigenvectors $\bar{\psi}_s^{(\mathbf{k})}$ ($s = \pm 1$) is set to be unity. For arbitrary $E_z^{(\mathbf{k})}$, we have $\psi_T^{(\mathbf{k})} = E_z^{(\mathbf{k})} \bar{\psi}_{\pm 1}^{(\mathbf{k})}$. From the Maxwell's equations it can be shown that the magnetic field $\mathbf{H}^{(\mathbf{k})}$ of the wavevector $\mathbf{k}$ satisfies

$H_x^{(\mathbf{k})} = s\dfrac{\sin\theta_{\mathbf{k}}}{v_g\tilde{\mu}} E_z^{(\mathbf{k})}$ and $H_y^{(\mathbf{k})} = -s\dfrac{\cos\theta_{\mathbf{k}}}{v_g\tilde{\mu}} E_z^{(\mathbf{k})}$, i.e., $\mathbf{H}^{(\mathbf{k})} = s\hat{\mathbf{k}}\times\mathbf{E}^{(\mathbf{k})}/(v_g\tilde{\mu})$ for $s = \pm 1$ (see Appendix A), where $\hat{\mathbf{k}} = \mathbf{k}/|\mathbf{k}|$. Thus, $s = +1$ and $-1$ describe the transverse modes in positive (right-handed) and negative (left-handed) media, respectively. In terms of electromagnetic fields, the eigenvectors become

$$\psi_T^{(\mathbf{k})} = \frac{1}{2}\begin{pmatrix} -\sqrt{\tilde{\mu}/\tilde{\varepsilon}}(iH_x^{(\mathbf{k})} + H_y^{(\mathbf{k})}) \\ \sqrt{2}E_z^{(\mathbf{k})} \\ \sqrt{\tilde{\mu}/\tilde{\varepsilon}}(iH_x^{(\mathbf{k})} - H_y^{(\mathbf{k})}) \end{pmatrix}. \tag{6}$$

Meanwhile, the eigenvector $\bar{\psi}_s^{(\mathbf{k})}$ ($s = 0$) corresponds to a longitudinal mode and can be written to the following form by the relation $\psi_L^{(\mathbf{k})} = \dfrac{\sqrt{2}H_x^{(\mathbf{k})}}{\cos\theta_{\mathbf{k}}}\bar{\psi}_0^{(\mathbf{k})}$ (see Appendix A),

$$\psi_L^{(\mathbf{k})} = \begin{pmatrix} H_x^{(\mathbf{k})} - iH_y^{(\mathbf{k})} \\ 0 \\ -H_x^{(\mathbf{k})} - iH_y^{(\mathbf{k})} \end{pmatrix}. \tag{7}$$

## B. Photonic analog of gate voltage

Potential, as a key quantity in quantum physics, is of central importance in the dynamics of particles. In graphene, the application of gate potentials can control electronic transport and give rise to various novel transport phenomena such as Klein tunneling[7,15] and supercollimation.[12-14] The interesting question is whether there exists a photonic analog of gate voltage in graphene and whether applications of such photonic potentials in PCs will bring us any novel transport behaviors.

Here we show that a photonic potential can be easily achieved by using the scaling properties of Maxwell's equations.[29] If the linear dimensions of the structures in a given dielectric PC are scaled uniformly by a factor of $\alpha$, the frequency $\omega$ and wavevector $\mathbf{k}$ should also be scaled according to the relations $\omega' = \omega/\alpha$ and $\mathbf{k}' = \mathbf{k}/\alpha$. Here we have assumed that the permittivity of the dielectric material of the 2D PC is independent of frequency in the frequency range of interest. These relations indicate that the scaling can lead to a shift of the Dirac-like point frequency, $\Delta\omega_D = \omega'_D - \omega_D = [1/\alpha - 1]\omega_D$, where $\omega_D$ and $\omega'_D$ are the Dirac-like point frequencies of the original and scaled PCs, respectively. It is easy to see that the group velocity is scaling invariant, i.e.,

$$\mathbf{v}'_g = \nabla_{\mathbf{k}'}\omega' = \nabla_{\alpha\mathbf{k}'}(\alpha\omega') = \nabla_{\mathbf{k}}\omega = \mathbf{v}_g, \tag{8}$$

where $\mathbf{v}_g$ and $\mathbf{v}'_g$ are the group velocities before and after the scaling, respectively. Thus, the variation of length scales of PCs mimicks the gate voltage in graphene, leading to a rigid shift of the Dirac-like cones in frequency. As a photonic analog of electron potential, the length-scaling induced Dirac-like point frequency shift ($\Delta\omega_D$) is effectively a photonic potential $V$, that is, $V = \Delta\omega_D = \omega'_D - \omega_D$. In order to construct a 1D photonic potential along one direction, say, $V(x) = \omega'_D(x) - \omega_D$, we can make the local length scales of the PCs a function of $x$ according to $V(x)$. The total Hamiltonian now becomes

$$H = v_g \mathbf{S} \cdot \mathbf{k} + V(x)\mathbf{I}, \tag{9}$$

where **I** is a 3-by-3 identity matrix. Due to the chiral nature of Eq. (9), the wave propagation along the *x*-direction produces no backscattering from the 1D potential and, therefore, gives rise to a one-way transport of EM waves. From an EM wave point of view, this is consistent with the fact that the impedance near the Dirac-like point frequency is a scaling invariant constant (see Appendix C).

## C. Super Klein tunneling

Using the effective photonic potential to control wave transport, we first study the scattering of pseudospin-1 EM waves in two dimensions by a square potential barrier along the *x*-direction as shown in Fig. 2(a). As predicted by previous work,[18-21] there exists a super Klein tunneling effect, i.e., unity transmission for all incident angles for pseudospin-1 EM waves when the frequency of the incident EM waves $\delta\omega = V_0/2$, where $V_0$ is the potential barrier height (see Appendix D). Such scattering properties can be traced to the boundary conditions given by the three-component spinor which are distinct from those of psedospin-½ (see Appendix E).

For an experimental implementation, the barrier can be realized by a sandwich structure composed of two PCs with different length scales. Figure 2(b) shows the schematic of one possible realization of such a barrier. Both photonic crystals labelled as PC1 and PC2 are square arrays of dielectric cylinders in air with dielectric constant $\varepsilon = 12.5$. The radii of two PCs are $r_i = 0.2a_i$ ($i = 1, 2$) with $a_1 = (15/14)a_2$, where $a_1$ and $a_2$ are the lattice constants of PC1 and PC2, respectively. Both PCs exhibit conical dispersions near $\mathbf{k} = 0$[24] (shown in Fig. 1) with the same group velocity $v_g = 0.2962c$ with $c$ being the speed of light in vacuum. The corresponding Dirac-like point frequencies are $\omega_{D1} = 1.0826\pi \frac{c}{a_1}$ and $\omega_{D2} = 1.0826\pi \frac{c}{a_2}$. It has been shown previously that such PCs can be described by effective medium theory near their respective Dirac-like point frequencies.[24] Thus, the photonic potential shift created by the presence of PC2 is $V_0 = \omega_{D2} - \omega_{D1} = \frac{1}{15}\omega_{D2}$. In Fig. 2(b), we set the thicknesses of PC1 and PC2 as $21a_1$ and $45a_2$, respectively. Then, we calculate the transmission of the PC sandwich structure as a function of incident angles at the reduced frequency $\delta\omega = \omega - \omega_{D1} = V_0/2$ using the commercial software package

COMSOL Multiphysics.[30] The result is shown in Fig. 3 (blue open circles). The nearly unity transmission holds for incident angles up to $80^0$, which agrees very well with the theoretical prediction of super Klein tunneling using the effective pseudospin-1 Hamiltonian[18-21] (red solid line). We note that the sharp transmission dip around $\theta \approx \pm 6.2^0$ for the PC sandwich structure comes from Rayleigh Wood anomalies[31] because PC1 and PC2 share a common spatial period in our simulation. This dip will not be present if the two PCs have incommensurate spatial periods. For higher angles, the effective medium approximation is no longer valid and thus the result from the real structure deviates from that of the effective Hamiltonian. The excellent agreement between the prediction from the effective Hamiltonian and the full-wave calculation validates the use of both the spin-orbit Hamiltonian and photonic potential in Eq. (9). To see the impact of the pseudospin number (1 vs. ½) on the scattering properties, in Fig. 3, we also plot the transmission for Dirac electrons in graphene (green open triangles). For a fair comparison, the electron energy $E$, potential height $U_0$ and width $D_e$ are taken as follows: $E/\hbar v_F = \delta\omega/v_g$, $U_0/\hbar v_F = V_o/v_g$ and $D_e = D = 45a_2$, where $v_F$ is the Fermi velocity of electrons in graphene. Different from the super Klein tunneling of pseudospin-1 EM waves at the reduced frequency $\delta\omega = V_0/2$, Dirac electrons of pseudospin-½ in graphene have unity transmission only at $\theta = 0^0$ and some finite angles satisfying the Fabry-Perot resonance conditions, indicating that Dirac electrons experience very strong anisotropic scattering at $E = U_0/2$ due to the 1D potential. The difference is due to different boundary conditions as explained analytically in Appendix F.

The super Klein tunneling of pseudospin-1 EM waves can also be understood from the concept of "complementary materials"[32] in EM wave theory. For example, in a slab composing of two equally thick adjacent layers, when the permittivity and permeability of one layer are opposite in sign to those of the other layer, these two layers will "optically cancel" each other in space.[32] A direct consequence of complementary materials is that the fields at input and output surfaces are identical,[32] which implies that their transfer matrix is an identity matrix for any incident angle. It is shown in Appendix G that the transfer matrix of an ABA structure of pseudospin-1 EM waves is exactly an identity matrix for all incident angles when $\delta\omega = V_0/2$, independent of the value of $V_0$. The pseudospin-1 photonic system is hence one possible platform to realize "complementary materials". It is worth noting that "complementary materials" are frequently discussed and studied within the context of metamaterials that involve complex metallic elements due to the requirement of negative refractive index. As such, the prospect of realizing the potential of this powerful concept at optical frequencies is not too promising due to the intrinsic absorption of metals. Pseudospin-1 photonic system provides a good alternative as they can be readily implemented in dielectric photonic crystals with low loss. We also note that the transfer matrix for pseudospin-½ electrons is *not* an identity matrix for oblique incidence unless additional conditions such as Fabry-Perot resonance are satisfied. This is why pseudospin-½ systems can only exhibit Klein tunneling but not "super" Klein tunneling.

## D. Supercollimation in a superlattice of PCs

Recently, supercollimation has been predicted in some special graphene superlattices.[13] The supercollimation is a result of anisotropic renormalization of the group velocity of the 2D chiral electrons in a 1D periodic potential. In order to investigate the supercollimation for pseudospin-1 EM waves, we consider here a superlattice composed of two PCs, e.g., PC1 and PC2 in Fig. 2(b), arranged periodically in the $x$-direction. For simplicity, we assume that these two PCs have the same thickness $d$ as shown schematically in Fig. 4(a). This structure forms a Kronig-Penney type of photonic potential along the $x$-direction with the potential height $V_0$, width $d$ and lattice constant $L$ ( $L=2d$ ), as shown in Fig. 4(b). The 2D dispersion relation for this superlattice can be obtained using the TMM method (see Appendix D), which gives

$$\cos 2k_x d = \cos q_{1x} d \cos q_{0x} d - \frac{\sin q_{1x} d \sin q_{0x} d}{2} \left[ \frac{(\delta\omega - V_0) q_{0x}}{\delta\omega q_{1x}} + \frac{\delta\omega q_{1x}}{(\delta\omega - V_0) q_{0x}} \right], \quad (10)$$

where $q_{0x}^2 + k_y^2 = \left(\delta\omega / v_g\right)^2$, $q_{1x}^2 + k_y^2 = \left[(\delta\omega - V_0)/v_g\right]^2$, $k_x$ is the Bloch wavevector in the Brillouin zone of the superlattice, and $k_y$ is the $y$ component of the wavevector. It is interesting to point out that when $\delta\omega = V_0/2$, equation (10) has a solution $k_x = 0$ for any $k_y$, i.e., the equifrequency contour is a straight line along the $k_y$ axis, indicating a zero group velocity in the $y$ direction. This is true independent of the value of $V_0$ ($V_0 \neq 0$). A typical dispersion relation of Eq. (10) is shown in Fig. 5(a) for the case of $V_0 = \frac{1}{15}\omega_{D2}$, which is the Dirac-like point frequency difference between PC1 and PC2, $d = 15a_2$ and $L = 30a_2$ ($a_2$ is the lattice constant of PC2). Figure 5(a) clearly exhibits a wedge-shaped structure, showing a dispersionless behavior along the $k_y$ direction even when $\delta\omega \neq V_0/2$. By expanding Eq. (10) around $\delta\omega = V_0/2$, we find a wedge equation,

$$\delta\omega = sv_g|k_x| + V_0/2 \ (s = \pm 1) \text{ when } |k_y| \ll \frac{V_0}{2v_g},$$

which shows explicitly that the group velocity along the $k_y$ direction vanishes while the one along $k_x$ direction remains unchanged within the range of the wedge-shaped structure. The strongly anisotropic renormalization of the group velocity shown in the wedge structure is the cause of supercollimation found in graphene.[12,13] Due to both the one-way transport along the *x*-direction and the zero group velocity in the *y*-direction it is expected that a wave packet constructed in the region of wedge should be guided to propagate undistorted along the periodic direction of the superlattice, independent of its initial direction of motion.

Here we should point out the difference between pseudospin-1 and pseudospin-½. For graphene, the wedge structure can only be realized in some specific periodic potentials $U_0$, i.e., $U_0 = 2\pi\hbar v_F/d$, and is limited to small values of $k_y$.[12-14] For pseudospin-1 EM waves, the existence of a wedge structure is much more robust. In fact, it appears for *any* finite value of $V_0$ as long as the two layers of the superlattice are of equal thickness, i.e., $L = 2d$. The wedge structure also exists in a much larger range in *k*-space. In Fig. 5(b), we show the dispersion relations for three fixed values of $k_y$. We find that the three dispersions are almost the same and agree very well with the wedge equation for the entire supercell Brillouin zone.

To demonstrate numerically the supercollimation of pseudospin-1 EM waves, we first connect the superlattice in Fig. 4(b) with $V_0 = \omega_{D2}/15$ and $d = 15a_2$ to a lead which is a PC with a photonic potential $V_0/2$, and then send a Gaussian wave packet from the lead towards the superlattice. At $t = 0$, we prepare a wave packet in the form $E_z = E_0 \exp\left[-|\mathbf{r}-\mathbf{r_c}|^2/r_0^2 + i\mathbf{k_c}\cdot(\mathbf{r}-\mathbf{r_c})\right]$, as shown in Fig. 6(a), with an initial center position of wave

packet $\mathbf{r_c}$ (a distance $d$ away from the superlattice), a reduced center frequency $\overline{\delta\omega_c} \equiv \delta\omega - V_0/2 = v_g |\mathbf{k_c}| = 0.06\pi v_g / L$ and a half width of $r_0 = 30d$. By using the TMM method (see Appendix D), we study the pulse propagation in the superlattice. The distributions of the electric field amplitude at $t = 1200d/v_g$ are shown in Figs. 6(b) and 6(c) for two different incident angles, $\theta = 0^0$ and $45^0$, respectively. In the absence of the superlattice, the wave packet propagates as a cylindrical wave along the direction of initial center wave vector and spreads out quickly. However, in the superlattice, the Gaussian wave packet is guided to propagate along the periodic direction ($x$-direction), regardless of the initial propagation direction. We notice that at the angle $\theta = 0^0$, the wave packet propagates in the superlattice with undistorted shape, but at the angle $\theta = 45^0$, the wave packet is stretched and tilted. This distortion of shape is induced by the strong $k_y$ dependent reflections around $\theta = 45^0$ at the interface between the lead and the superlattice. After the wave packet has entered the superlattice, our calculations show that its shape remains unchanged as expected.

## III. CONCLUSIONS

In summary, we have demonstrated that certain dielectric photonic crystals exhibiting Dirac-like conical dispersion at $\mathbf{k} = 0$ can be used to realize photonic pseudospin-1 materials and found that the length-scaling induced frequency shift in the PCs can be mapped to an effective photonic potential. With the introduction of the effective photonic potential, we have also demonstrated the super Klein tunneling effect for pseudospin-1 EM waves. In a superlattice of PCs, we have found much more robust supercollimation for pseudospin-1 EM waves than pseudospin-½ electrons due to a wedge structure in the dispersion which holds for any finite value of potential height $V_0$. Both the super Klein tunneling and robust supercollimation can be understood from the perspective of complementary materials in EM wave theory. It should be mentioned that the results obtained here also apply to the transverse magnetic (TM) waves if we simply switch the electric field with magnetic field and electric permittivity with magnetic permeability. Compared to the proposals of constructing artificial crystals in ultracold atom systems, our implementation using dielectric PCs greatly reduces technical difficulties in experiments and paves the way to realistic observations of the pseudospin-1 physics, including but not limited to super Klein tunneling and supercollimation.

# ACKNOWLEDGMENTS

This work is supported by Research Grants Council, University Grants Committee, Hong Kong (AoE/P-02/12).

# APPENDIX A

For TE modes, the full set of Maxwell's equations reduces to

$$\frac{\partial E_z}{\partial y} = i\omega\mu H_x, \tag{A1}$$

$$\frac{\partial E_z}{\partial x} = -i\omega\mu H_y, \tag{A2}$$

$$\frac{\partial H_y}{\partial x} - \frac{\partial H_x}{\partial y} = -i\omega\varepsilon E_z, \tag{A3}$$

$$\frac{\partial H_x}{\partial x} + \frac{\partial H_y}{\partial y} = 0, \tag{A4}$$

for transverse modes and

$$\frac{\partial H_y}{\partial x} - \frac{\partial H_x}{\partial y} = 0, \tag{A5}$$

for longitudinal modes with $E_z$ being spatially independent. Here $\mu = \mu_{xx} = \mu_{yy}$ and $\varepsilon = \varepsilon_{zz}$. Note that the longitudinal modes only exist at the frequency where both effective parameters vanish ($\varepsilon = \mu = 0$). By combining Eqs. (A1)-(A5), we can obtain Eq. (1) in the text.

With the Fourier transforms $H_m^{(\mathbf{k})} = \frac{1}{2\pi}\int H_m e^{-i\mathbf{k}\cdot\mathbf{r}} d\mathbf{r}$ ($m = x, y$) and $E_z^{(\mathbf{k})} = \frac{1}{2\pi}\int E_z e^{-i\mathbf{k}\cdot\mathbf{r}} d\mathbf{r}$, the Maxwell's equations for transverse modes, i.e., equations (A1) and (A2), can be written as $H_x^{(\mathbf{k})} = \frac{k_y}{\omega\mu} E_z^{(\mathbf{k})}$ and $H_y^{(\mathbf{k})} = -\frac{k_x}{\omega\mu} E_z^{(\mathbf{k})}$, where $k_x$ and $k_y$ are the $x$ and $y$ components of the 2D wavevector $\mathbf{k}$ in the $xy$ plane, respectively. Near a Dirac-like point, $\omega\mu$ can be approximated as $\omega\mu \cong \omega_D(\omega - \omega_D)\frac{d\mu}{d\omega}\bigg|_{\omega=\omega_D} = (\omega - \omega_D)\tilde{\mu}$.

By using the linear dispersion $\omega - \omega_D = sv_g|\mathbf{k}|$, we obtain

$$H_x^{(\mathbf{k})} \cong \frac{k_y}{(\omega-\omega_D)\tilde{\mu}} E_z^{(\mathbf{k})} = \frac{k_y}{sv_g |\mathbf{k}|\tilde{\mu}} E_z^{(\mathbf{k})} = s\frac{\sin\theta_\mathbf{k}}{v_g \tilde{\mu}}, \tag{A6}$$

and

$$H_y^{(\mathbf{k})} \cong -\frac{k_x}{(\omega-\omega_D)\tilde{\mu}} E_z^{(\mathbf{k})} = -\frac{k_x}{sv_g |\mathbf{k}|\tilde{\mu}} E_z^{(\mathbf{k})} = -s\frac{\cos\theta_\mathbf{k}}{v_g \tilde{\mu}}, \tag{A7}$$

where $s = \mathrm{sgn}(\omega-\omega_D)$ and $\theta_\mathbf{k}$ is the angle of the wavevector $\mathbf{k}$ with respect to the x-axis, i.e., $k_x = |\mathbf{k}|\cos\theta_\mathbf{k}$, $k_y = |\mathbf{k}|\sin\theta_\mathbf{k}$. Thus, the eigenvectors of Eq. (4) for arbitrary $E_z^{(\mathbf{k})}$, $\psi_T^{(\mathbf{k})} = E_z^{(\mathbf{k})}\overline{\psi}_{\pm 1}^{(\mathbf{k})}$, can be written in the form of Eq. (6). For longitudinal modes, equation (A5) can be written as $H_x^{(\mathbf{k})}k_y = H_y^{(\mathbf{k})}k_x$, i.e., $H_y^{(\mathbf{k})}/H_x^{(\mathbf{k})} = \sin\theta_\mathbf{k}/\cos\theta_\mathbf{k}$, and Eq. (7) can be obtained from $\psi_L^{(\mathbf{k})} = \frac{\sqrt{2}H_x^{(\mathbf{k})}}{\cos\theta_\mathbf{k}}\overline{\psi}_0^{(\mathbf{k})}$.

## APPENDIX B

Let $\tilde{\psi}^{(\mathbf{k})}$ be the Fourier transform of $\tilde{\psi}$ in k-space, we have $\tilde{\psi}^{(\mathbf{k})} = \frac{1}{2\pi}\int \tilde{\psi}e^{-i\mathbf{k}\cdot\mathbf{r}}d\mathbf{r}$, i.e.,

$$\tilde{\psi}_T^{(\mathbf{k})} = \frac{1}{2\pi}\int \tilde{\psi}_T e^{-i\mathbf{k}\cdot\mathbf{r}}d\mathbf{r} = (-iH_x^{(\mathbf{k})} - H_y^{(\mathbf{k})}, E_z^{(\mathbf{k})}, iH_x^{(\mathbf{k})} - H_y^{(\mathbf{k})})^T \text{ for transverse modes and}$$

$$\tilde{\psi}_L^{(\mathbf{k})} = \frac{1}{2\pi}\int \tilde{\psi}_L e^{-i\mathbf{k}\cdot\mathbf{r}}d\mathbf{r} = (H_x^{(\mathbf{k})} - iH_y^{(\mathbf{k})}, E_z^{(\mathbf{k})}, -H_x^{(\mathbf{k})} - iH_y^{(\mathbf{k})})^T \text{ for longitudinal modes, with}$$

$H_m^{(\mathbf{k})} = \frac{1}{2\pi}\int H_m e^{-i\mathbf{k}\cdot\mathbf{r}}d\mathbf{r}$ ($m = x, y$), and $E_z^{(\mathbf{k})} = \frac{1}{2\pi}\int E_z e^{-i\mathbf{k}\cdot\mathbf{r}}d\mathbf{r}$. Thus, with the approximations of

$\omega\varepsilon \cong \omega_D \left.\frac{d\varepsilon}{d\omega}\right|_{\omega=\omega_D}(\omega-\omega_D) \equiv (\omega-\omega_D)\tilde{\varepsilon}$ and $\omega\mu \cong \omega_D \left.\frac{d\mu}{d\omega}\right|_{\omega=\omega_D}(\omega-\omega_D) \equiv (\omega-\omega_D)\tilde{\mu}$ near the Dirac-

like point frequency, we can rewrite Eq. (1) in k-space as

$$\begin{pmatrix} 0 & k_x - ik_y & 0 \\ k_x + ik_y & 0 & k_x - ik_y \\ 0 & k_x + ik_y & 0 \end{pmatrix}\tilde{\psi}^{(\mathbf{k})} = \delta\omega \begin{pmatrix} \tilde{\mu} & 0 & 0 \\ 0 & 2\tilde{\varepsilon} & 0 \\ 0 & 0 & \tilde{\mu} \end{pmatrix}\tilde{\psi}^{(\mathbf{k})}, \tag{B1}$$

where $\delta\omega = \omega - \omega_D$. Here $k_x$ and $k_y$ are the *x* and *y* components of the 2D wavevector **k** in the *xy* plane, respectively. It should be mentioned that the component $E_z^{(\mathbf{k})}$ of longitudinal modes is zero for any non-zero **k**. Equation (B1) can be transformed into a simple eigenvalue problem if we left multiply both sides of Eq. (B1) by a matrix $U$ and write $\tilde{\psi}^{(\mathbf{k})} = U\psi^{(\mathbf{k})}$, i.e.,

$$\frac{1}{\sqrt{2\tilde{\varepsilon}\tilde{\mu}}}\begin{pmatrix} 0 & k_x - ik_y & 0 \\ k_x + ik_y & 0 & k_x - ik_y \\ 0 & k_x + ik_y & 0 \end{pmatrix}\psi^{(\mathbf{k})} = \delta\omega\psi^{(\mathbf{k})}, \tag{B2}$$

with

$$U = \begin{pmatrix} 2\sqrt{\tilde{\varepsilon}/\tilde{\mu}} & 0 & 0 \\ 0 & \sqrt{2} & 0 \\ 0 & 0 & 2\sqrt{\tilde{\varepsilon}/\tilde{\mu}} \end{pmatrix}. \tag{B3}$$

## APPENDIX C

From the effective medium theory,[33] the effective permittivity $\varepsilon$ and permeability $\mu$ are functions of the dimensionless frequency, $\bar{\omega} = \omega a / 2\pi c$, where $a$ is the lattice constant of the PC and $c$ is the speed of light in vacuum. For PCs with a Dirac-like cone, the effective permittivity $\varepsilon$ and permeability $\mu$ vanish simultaneously at the Dirac-like point frequency, $\bar{\omega}_D$, i.e., $\varepsilon(\bar{\omega}_D) = \mu(\bar{\omega}_D) = 0$. Then the permittivity $\varepsilon$ and permeability $\mu$ can be approximated as $\varepsilon \cong (\bar{\omega} - \bar{\omega}_D)\frac{d\varepsilon}{d\bar{\omega}}\bigg|_{\bar{\omega}=\bar{\omega}_D}$ and $\mu \cong (\bar{\omega} - \bar{\omega}_D)\frac{d\mu}{d\bar{\omega}}\bigg|_{\bar{\omega}=\bar{\omega}_D}$. We can obtain the impedance $Z$ near the Dirac-like point frequency,

$$Z(\bar{\omega}) = \sqrt{\mu(\bar{\omega})/\varepsilon(\bar{\omega})} \cong \sqrt{\frac{(\bar{\omega}-\bar{\omega}_D)\frac{d\mu}{d\bar{\omega}}\bigg|_{\bar{\omega}=\bar{\omega}_D}}{(\bar{\omega}-\bar{\omega}_D)\frac{d\varepsilon}{d\bar{\omega}}\bigg|_{\bar{\omega}=\bar{\omega}_D}}} = \sqrt{\frac{\tilde{\mu}}{\tilde{\varepsilon}}}, \tag{C1}$$

where $\tilde{\mu} = \bar{\omega}_D \left.\dfrac{d\mu}{d\bar{\omega}}\right|_{\bar{\omega}=\bar{\omega}_D} = \omega_D \left.\dfrac{d\mu}{d\omega}\right|_{\omega=\omega_D}$ and $\tilde{\varepsilon} = \bar{\omega}_D \left.\dfrac{d\varepsilon}{d\bar{\omega}}\right|_{\bar{\omega}=\bar{\omega}_D} = \omega_D \left.\dfrac{d\varepsilon}{d\omega}\right|_{\omega=\omega_D}$. Due to the scaling properties of Maxwell's equations, the dimensionless Dirac-like point frequency $\bar{\omega}_D$ is a scaling invariant constant. Thus, the impedance near the Dirac-like point frequency is a scaling invariant constant as well as $\tilde{\varepsilon}$ and $\tilde{\mu}$.

For an effective homogeneous medium, the angular frequency $\omega$, the refractive index $n_p$ and the wavenumber $k$ are related by $\omega = \dfrac{kc}{n_p}$. Thus, the group velocity $v_g$ is written as

$$v_g = \dfrac{d\omega}{dk} = \dfrac{c}{n_p} - \dfrac{kc}{n_p^2}\dfrac{dn_p}{d\omega}\dfrac{d\omega}{dk} = \dfrac{c}{n_p} - \dfrac{\omega}{n_p}\dfrac{dn_p}{d\omega}v_g, \text{ i.e., } v_g = \dfrac{c}{n_p + \omega\dfrac{dn_p}{d\omega}}.$$

With the relation $\bar{\omega} = \omega a/2\pi c$, we obtain $v_g = \dfrac{c}{n_p + \bar{\omega}\dfrac{dn_p}{d\bar{\omega}}}$. For a PC with a Dirac-like cone, the effective refractive index in a small region close to the Dirac-like point frequency, i.e., $\delta\omega/\omega_D \ll 1$, can be approximated as

$$n_p = c\sqrt{\varepsilon(\bar{\omega})}\sqrt{\mu(\bar{\omega})} \cong c(\bar{\omega}-\bar{\omega}_D)\sqrt{\left.\dfrac{d\varepsilon}{d\bar{\omega}}\right|_{\bar{\omega}=\bar{\omega}_D}\left.\dfrac{d\mu}{d\bar{\omega}}\right|_{\bar{\omega}=\bar{\omega}_D}}.$$

Hence its group velocity becomes

$$v_g = \dfrac{c}{n_p + \bar{\omega}\dfrac{dn_p}{d\bar{\omega}}} \cong \dfrac{c}{c(\bar{\omega}-\bar{\omega}_D)+c\bar{\omega}}\dfrac{1}{\sqrt{\left.\dfrac{d\varepsilon}{d\bar{\omega}}\right|_{\bar{\omega}=\bar{\omega}_D}\left.\dfrac{d\mu}{d\bar{\omega}}\right|_{\bar{\omega}=\bar{\omega}_D}}} \cong \dfrac{1}{\sqrt{\tilde{\varepsilon}\tilde{\mu}}},$$

which is non-zero even though the phase velocity approaches infinity.

## APPENDIX D

For pseudospin-1 EM waves transmitted through a square potential barrier as shown in Fig. 2(a), the wave function in different regions can be written in terms of the incident and reflected waves with the eigenvectors in Eq. (5). In region I, we have

$$\psi_{\mathrm{I}} = \frac{\tilde{a}_0}{2} \begin{pmatrix} se^{-i\theta} \\ \sqrt{2} \\ se^{i\theta} \end{pmatrix} e^{i(q_{0x}x+q_{0y}y)} + \frac{\tilde{b}_0}{2} \begin{pmatrix} se^{-i(\pi-\theta)} \\ \sqrt{2} \\ se^{i(\pi-\theta)} \end{pmatrix} e^{i(-q_{0x}x+q_{0y}y)}, \tag{D1}$$

with $\theta$ as the angle of the wavevector $\mathbf{q_0}$ with respect to the $x$-axis, $s = \mathrm{sgn}(\delta\omega)$, $q_{0x} = |\mathbf{q_0}|\cos\theta$, $q_{0y} = |\mathbf{q_0}|\sin\theta = k_y$ and $|\mathbf{q_0}| = |\delta\omega|/v_g$. In region II, we have

$$\psi_{\mathrm{II}} = \frac{\tilde{c}}{2} \begin{pmatrix} s'e^{-i\phi} \\ \sqrt{2} \\ s'e^{i\phi} \end{pmatrix} e^{i(q_{1x}x+q_{1y}y)} + \frac{\tilde{d}}{2} \begin{pmatrix} s'e^{-i(\pi-\phi)} \\ \sqrt{2} \\ s'e^{i(\pi-\phi)} \end{pmatrix} e^{i(-q_{1x}x+q_{1y}y)}, \tag{D2}$$

with $\phi$ as the angle of the wavevector $\mathbf{q_1}$ with respect to the $x$-axis, $s' = \mathrm{sgn}(\delta\omega - V_0)$, $q_{1x} = |\mathbf{q_1}|\cos\phi$, $q_{1y} = |\mathbf{q_1}|\sin\phi = |\mathbf{q_0}|\sin\theta = k_y$ and $|\mathbf{q_1}| = |\delta\omega - V_0|/v_g$. And in region III, we have

$$\psi_{\mathrm{III}} = \frac{\tilde{a}_1}{2} \begin{pmatrix} se^{-i\theta} \\ \sqrt{2} \\ se^{i\theta} \end{pmatrix} e^{i(q_{0x}x+q_{0y}y)} + \frac{\tilde{b}_1}{2} \begin{pmatrix} se^{-i(\pi-\theta)} \\ \sqrt{2} \\ se^{i(\pi-\theta)} \end{pmatrix} e^{i(-q_{0x}x+q_{0y}y)}. \tag{D3}$$

We define the transfer matrix $\mathbf{M}$ by the relation,

$$\begin{pmatrix} a_1 \\ b_1 \end{pmatrix} = \mathbf{M} \begin{pmatrix} a_0 \\ b_0 \end{pmatrix}, \tag{D4}$$

where $a_0 = \tilde{a}_0 e^{iq_{0x}x_0}$, $b_0 = \tilde{b}_0 e^{-iq_{0x}x_0}$, $a_1 = \tilde{a}_1 e^{iq_{0x}(x_0+D)}$ and $b_1 = \tilde{b}_1 e^{-iq_{0x}(x_0+D)}$ ($x = x_0$ and $x = x_0 + D$ are the two boundaries of the square barrier). From the continuity of $\psi_2$ and $\psi_1 + \psi_3$ at the boundaries (see Appendix E for boundary conditions), the transfer matrix can be obtained and has the form,

$$\mathbf{M}(D) = \begin{pmatrix} \alpha(D) & \beta(-D) \\ \beta(D) & \alpha(-D) \end{pmatrix}, \tag{D5}$$

with the elements,

$$\alpha(D) = \cos q_{1x}D + \frac{i}{2} ss' \sin q_{1x}D \left( \frac{\cos\theta}{\cos\phi} + \frac{\cos\phi}{\cos\theta} \right),$$

$$\beta(D) = \frac{i}{2} ss' \sin q_{1x}D \left( \frac{\cos\theta}{\cos\phi} - \frac{\cos\phi}{\cos\theta} \right).$$

(D6)

For Klein tunneling, since there is no reflected wave in region III, we have $b_1 = 0$. Then the transmission through the barrier is

$$T = \frac{\cos^2\phi \cos^2\theta}{\left(\cos q_{1x}D \cos\phi \cos\theta\right)^2 + \frac{\sin^2 q_{1x}D}{4}\left(\cos^2\phi + \cos^2\theta\right)^2}.$$

(D7)

When $\delta\omega = V_0/2$, we have $|\mathbf{q}_1| = |\mathbf{q}_0|$ and $\phi = \pi - \theta$. Thus, we obtain $T = 1$ for any $\theta$ value from Eq. (D7), i.e., the super Klein tunneling effect.

For a Kronig-Penney type of photonic potential with the lattice constant $L$, barrier width $d$ and barrier height $V_0$, as shown in Fig. 4(b), the transfer matrix for one unit cell is

$$\mathbf{M}_{cell} = \mathbf{P}(L-d)\mathbf{M}(d),$$

(D8)

with the matrix,

$$\mathbf{P}(L-d) = \begin{pmatrix} e^{iq_{0x}(L-d)} & 0 \\ 0 & e^{-iq_{0x}(L-d)} \end{pmatrix}.$$

(D9)

From the Bloch theorem, we have the periodic boundary condition, $\psi(x=L) = e^{ik_x L}\psi(x=0)$ ($k_x$ is the Bloch wavevector in the supercell Brillouin zone), i.e.,

$$e^{ik_x L}\begin{pmatrix} a_0 \\ b_0 \end{pmatrix} = \mathbf{M}_{cell}\begin{pmatrix} a_0 \\ b_0 \end{pmatrix}.$$

(D10)

Thus, we can solve the secular equation, $\left|\mathbf{M}_{cell} - e^{ik_x L}\mathbf{I}\right| = 0$, to obtain the dispersion relation,

$$\cos k_x L = \cos q_{1x} d \cos q_{0x}(L-d) - \frac{ss'}{2} \sin q_{1x} d \sin q_{0x}(L-d)\left(\frac{\cos\phi}{\cos\theta} + \frac{\cos\theta}{\cos\phi}\right), \tag{D11}$$

which leads to Eq. (10) in the text when $L = 2d$.

# APPENDIX E

To determine the propagation of pseudospin-1 EM waves in the 1D potential, it is necessary to obtain the boundary conditions at the interface. For a general solution, $\psi = (\psi_1, \psi_2, \psi_3)^T$ of Eq. (9), i.e., $H\psi = \delta\omega\psi$, we have

$$\frac{v_g}{\sqrt{2}}\left(-i\frac{\partial\psi_2}{\partial x} - \frac{\partial\psi_2}{\partial y}\right) + V(x)\psi_1 = \delta\omega\psi_1, \tag{E1}$$

$$\frac{v_g}{\sqrt{2}}\left(-i\frac{\partial\psi_1}{\partial x} + \frac{\partial\psi_1}{\partial y} - i\frac{\partial\psi_3}{\partial x} - \frac{\partial\psi_3}{\partial y}\right) + V(x)\psi_2 = \delta\omega\psi_2, \tag{E2}$$

$$\frac{v_g}{\sqrt{2}}\left(-i\frac{\partial\psi_2}{\partial x} + \frac{\partial\psi_2}{\partial y}\right) + V(x)\psi_3 = \delta\omega\psi_3. \tag{E3}$$

For an interface at $x = x_0$, we integrate Eq. (E1) from $x_0 - \epsilon$ to $x_0 + \epsilon$ and then take the limit $\epsilon \to 0$:

$$-i\frac{v_g}{\sqrt{2}}\int_{x_0-\epsilon}^{x_0+\epsilon}\frac{\partial\psi_2}{\partial x}dx - \frac{v_g}{\sqrt{2}}\int_{x_0-\epsilon}^{x_0+\epsilon}\frac{\partial\psi_2}{\partial y}dx + \int_{x_0-\epsilon}^{x_0+\epsilon}V(x)\psi_1 dx = \delta\omega\int_{x_0-\epsilon}^{x_0+\epsilon}\psi_1 dx. \tag{E4}$$

The last three integrals are zero in the limit $\epsilon \to 0$ with the assumption that $V(x)$ and the wave function components are all finite. Then we obtain the first boundary condition from the first integral, $\psi_2(x_0 - \epsilon) = \psi_2(x_0 + \epsilon)$, i.e., $\psi_2$ is continuous. Similarly, integrating Eq. (E2) gives us the second boundary condition, $\psi_1(x_0 - \epsilon) + \psi_3(x_0 - \epsilon) = \psi_1(x_0 + \epsilon) + \psi_3(x_0 + \epsilon)$, i.e., $\psi_1 + \psi_3$ is continuous. Subtracting Eq. (E1) from Eq. (E3), we have $\sqrt{2}v_g\frac{\partial\psi_2}{\partial y} = [\delta\omega - V(x)](\psi_3 - \psi_1)$. Due to the continuity of $\psi_2$ at the interface $x = x_0$, the derivative of $\psi_2$ along the direction parallel to the boundary, $\dfrac{\partial \psi_2}{\partial y}$, is also continuous, leading to the third boundary condition that $[\delta\omega - V(x)](\psi_3 - \psi_1)$ is continuous. Notice that the effective impedance $Z = \sqrt{\tilde{\mu}/\tilde{\varepsilon}}$ and the parameter $\tilde{\mu}$ are scaling invariant constants (see Appendix C) and $\omega\mu \cong (\omega - \omega_D)\tilde{\mu} = [\delta\omega - V(x)]\tilde{\mu}$. Thus, by applying these boundary conditions to the wave functions in Eqs. (6) and (7), we obtain the continuity of $E_z$, $H_y$ and $\mu H_x$, respectively, which are the same boundary conditions as required by TE waves.

It should be noted that among the above three boundary conditions, only the first and second boundary conditions are independent since the continuity of $\psi_2(x_0)$ implies the continuity of $[\delta\omega - V(x_0)][\psi_3(x_0) - \psi_1(x_0)]$. It is consistent with EM wave theory where the continuity of normal **B** and **D** components across interfaces are implied by the continuity of the tangential **E** and **H** components for time-harmonic fields.[34] It should be pointed out that unlike electrons in graphene, pseudospin-1 EM waves do not require each component of the wave function to be continuous at the boundaries. As shown in the text, this important difference of pseudospin-1 EM waves can lead to very different transport properties from those of pseudospin-½ graphene.

## APPENDIX F

Consider the incidence of Dirac electrons with energy $E$ from the left of a constant potential $U_0$ with an interface at $x = 0$. The wave functions in the regions $x < 0$ and $x > 0$ are, respectively,

$$\psi_1 = \begin{pmatrix} 1 \\ se^{i\theta} \end{pmatrix} e^{iq_{0x}x + ik_y y} + r_e \begin{pmatrix} 1 \\ se^{i(\pi-\theta)} \end{pmatrix} e^{-iq_{0x}x + ik_y y}, \tag{F1}$$

with $q_{0x} = |\mathbf{q_0}|\cos\theta$, $k_y = |\mathbf{q_0}|\sin\theta$, $|\mathbf{q_0}| = |E|/\hbar v_F$ and $s = \text{sgn}(E)$, and

$$\psi_2 = t_e \begin{pmatrix} 1 \\ s'e^{i\phi} \end{pmatrix} e^{iq_{1x}x + ik_y y}, \tag{F2}$$

with $q_{1x} = |\mathbf{q_1}|\cos\phi$, $|\mathbf{q_1}|\sin\phi = |\mathbf{q_0}|\sin\theta = k_y$, $|\mathbf{q_1}| = |E - U_0|/\hbar v_F$ and $s' = \text{sgn}(E - U_0)$. Here $r_e$ and $t_e$ are the reflection and transmission amplitudes, respectively.

From the continuity of each component of the spinor, we obtain $r_e = \dfrac{se^{i\theta} - s'e^{i\phi}}{se^{-i\theta} + s'e^{i\phi}}$ and the reflection coefficient,

$$R_e = |r_e|^2 = \frac{(s\cos\theta - s'\cos\phi)^2 + (s\sin\theta - s'\sin\phi)^2}{(s\cos\theta + s'\cos\phi)^2 + (s\sin\theta - s'\sin\phi)^2}. \tag{F3}$$

For comparison, we also study the reflection of pseudospin-1 EM waves with a reduced frequency $\delta\omega$ ($\delta\omega/v_g = E/\hbar v_F$) entering a constant potential $V_0$ with $V_0/v_g = U_0/\hbar v_F$. Due to different boundary conditions for pseudospin-1 EM waves discussed in Appendix E, we obtain a different reflection amplitude $r_p = \dfrac{s\cos\theta - s'\cos\phi}{s\cos\theta + s'\cos\phi}$. The reflection coefficient becomes

$$R_p = |r_p|^2 = \frac{(s\cos\theta - s'\cos\phi)^2}{(s\cos\theta + s'\cos\phi)^2}. \tag{F4}$$

Comparing Eqs. (F3) and (F4), it is easy to see that $R_e \geq R_p$ because both the numerator and denominator of $R_e$ can be obtained from those of $R_p$ by adding the same non-negative term $(s\sin\theta - s'\sin\phi)^2$. Especially, when $\delta\omega = V_0/2$ and $E = U_0/2$, we have $s' = -s$ and $\phi = \pi - \theta$, leading to $R_p = 0$ and $R_e = \sin^2\theta$. This difference is manifested in their corresponding Klein tunneling effects shown in the text.

## APPENDIX G

For pseudospin-1 EM waves with a frequency $\delta\omega > 0$ going through an ABA structure where slabs A and B have the thicknesses $d/2$ and $d$ and the photonic potentials $0$ and $V_0$, respectively, the system's transfer matrix can be obtained from

$$\mathbf{M}_{sys}^P = \mathbf{M}_A^P(d/2)\mathbf{M}_B^P(d)\mathbf{M}_A^P(d/2), \tag{G1}$$

where $\mathbf{M}_A^P$ and $\mathbf{M}_B^P$ are the respective transfer matrices of slabs A and B and have the forms (see Appendix D),

$$\mathbf{M}_A^P = \begin{pmatrix} e^{iq_{Ax}d/2} & 0 \\ 0 & e^{-iq_{Ax}d/2} \end{pmatrix}, \tag{G2}$$

$$\mathbf{M}_B^P = \begin{pmatrix} \alpha^P(d) & \beta^P(-d) \\ \beta^P(d) & \alpha^P(-d) \end{pmatrix}, \tag{G3}$$

with the elements,

$$\alpha^P(d) = \cos q_{Bx}d + \frac{i}{2} ss_B \sin q_{Bx}d \left( \frac{\cos\theta}{\cos\phi_B} + \frac{\cos\phi_B}{\cos\theta} \right),$$

$$\beta^P(d) = \frac{i}{2} ss_B \sin q_{Bx}d \left( \frac{\cos\theta}{\cos\phi_B} - \frac{\cos\phi_B}{\cos\theta} \right), \tag{G4}$$

where $\theta$ is the incident angle, $\phi_B = \sin^{-1}\left[|\delta\omega/(\delta\omega-V_0)|\sin\theta\right]$, $q_{Ax} = \cos\theta|\delta\omega|/v_g$ and $q_{Bx} = \cos\phi_B|\delta\omega-V_0|/v_g$. Here $s = \text{sgn}(\delta\omega)$ and $s_B = \text{sgn}(\delta\omega-V_0)$. When $\delta\omega = V_0/2$, we have $\phi_B = \pi - \theta$, $q_{Bx} = -q_{Ax}$ and $s_B = -s$. Applying these relations to Eq. (G1), we can obtain $\mathbf{M}_{sys}^P = \mathbf{I}$, where $\mathbf{I}$ is a 2-by-2 identity matrix, independent of the incident angle $\theta$. This result shows that the ABA sandwich structure acts as complementary materials for pseudospin-1 EM waves at $\delta\omega = V_0/2$.

Using the TMM method and taking steps similar to pseudospin-1 EM waves (see Appendix D), we can also obtain the transfer matrices of Dirac electrons with energy $E$ ($E/\hbar v_F = \delta\omega/v_g$) going through an ABA structure where slabs A and B have the respective potential heights 0 and $U_0$ ($U_0/\hbar v_F = V_0/v_g$) and slab widths $d/2$ and $d$,

$$\mathbf{M}_A^e = \begin{pmatrix} e^{iq_{Ax}d/2} & 0 \\ 0 & e^{-iq_{Ax}d/2} \end{pmatrix}, \tag{G5}$$

$$\mathbf{M}_B^e = \begin{pmatrix} \alpha^e(d) & \beta^{e*}(d) \\ \beta^e(d) & \alpha^{e*}(d) \end{pmatrix}, \tag{G6}$$

with the elements,

$$\alpha^e(d) = \cos q_{Bx}d + i\sin q_{Bx}d\left(ss_B \sec\theta \sec\phi_B - \tan\theta \tan\phi_B\right),$$
$$\beta^e(d) = e^{i\theta}\sin q_{Bx}d(\sec\theta \tan\phi_B - ss_B \sec\phi_B \tan\theta). \tag{G7}$$

Then, we study the transfer matrix of Dirac electrons for the sandwich system. Similarly, when $E = U_0/2$, we have $\phi_B = \pi - \theta$, $q_{Bx} = -q_{Ax} = -q_x$ and $s_B = -s$. The transfer matrix for the system is

$$\mathbf{M}^e_{sys} = \begin{pmatrix} \alpha^e(d)e^{iq_x d} & \beta^{e*}(d) \\ \beta^e(d) & \alpha^{e*}(d)e^{-iq_x d} \end{pmatrix}. \tag{G8}$$

Thus, we have $\mathbf{M}^e_{sys} = \mathbf{I}$ only for $\theta = 0$ or incident angles satisfying $q_{Bx}d = n\pi$ with $n$ being an integer. For other incident angles, $\mathbf{M}^e_{sys} \neq \mathbf{I}$. This indicates that pseudospin-½ electrons cannot have the same "complementary" property as pseudospin-1 EM waves.

# Figure Captions

FIG. 1. The band dispersion of a 2D photonic crystal (PC) with dielectric cylinders having radius $r = 0.2\text{a}$ and $\varepsilon = 12.5$ arranged in a square lattice in air (see inset). (a) The band structure with a region of linear dispersions intercepted by a flat band near the $\Gamma$ point (**k**=0) marked by the dashed rectangle. (b) The 3D band surfaces for the linear region marked in (a) exhibiting a conical dispersion near **k**=0.

FIG. 2. Klein tunneling of pseudospin-1 EM waves. (a) Schematic diagram of the scattering of psuedospin-1 EM waves by a square photonic potential barrier and definitions of the wavevectors $\mathbf{q_0}$ and $\mathbf{q_1}$ with their respective angles $\theta$ and $\phi$. (b) The realization of a square photonic potential barrier by a PC sandwich structure. The two PCs, PC1 and PC2, have the same structure as shown in Fig. 1, but differ by a length scale ($a_1 = 15a_2/14$ and $r_1 = 15r_2/14$). The thicknesses of PC1 and PC2 are $21a_1$ and $45a_2$, respectively. The change in length scale shifts the "Dirac-like point" while preserving the conical dispersion. Length-scale change in pseudospin-1 photonic system is hence an analog of potential change (gating) in pseudospin-½ graphene.

FIG. 3. Klein tunneling for pseudospin-1 EM waves and pseudospin-½ electrons. Transmission amplitudes are plotted for EM waves in a PC sandwich structure (blue open circles), in an effective spin-orbit Hamiltonian of Eq. (9) (red solid line), and electrons in graphene (green open triangles). The geometric parameters of the sandwich structure are shown in Fig. 2(b). The reduced frequency, photonic potential and group velocity in the effective Hamiltonian are chosen as $\delta\omega = V_0/2$, $V_0 = \omega_{D2}/15$ and $v_g = 0.2962c$. For graphene, the electron energy $E$ and potential $U_0$ are taken as $E/\hbar v_F = \delta\omega/v_g$ and $U_0/\hbar v_F = V_0/v_g$, respectively, where $v_F$ is the Fermi velocity of electrons in graphene. Both pseudospin-1 EM waves and pseudospin-½ electrons have a barrier width $45a_2$.

FIG. 4. Schematic of a superlattice of PCs. (a) A superlattice realized by stacking alternate layers of two PCs with different length scales. The layers have equal thickness $d$ and the spatial period $L = 2d$. (b) The pseudospin-1 photonic states feel a Kronig-Penney type of photonic potential formed by the superlattice in (a). The barrier height $V_0$ is the difference of the Dirac-like point frequencies of two PCs.

FIG. 5. Photonic dispersion of pseudospin-1 EM waves in a superlattice (Kronig-Penney photonic potential). (a) Band dispersion near $\delta\omega = V_0/2$ for the superlattice realized by layers of PC1 and PC2 with equal thickness $d = 15a_2$ ($a_2$ is the lattice constant of PC2) and the spatial period $L = 30a_2$. The "photonic potential" is shown in Fig. 4(b), with $V_0 = \omega_{D2}/15$. (b) Band dispersion relation, $\delta\omega$ vs $k_x$, for different values of $k_y$. Red solid lines, blue open circles and green solid triangles correspond to $k_y = 0$, $0.48\pi/L$, and $0.96\pi/L$, respectively.

FIG. 6. Supercollimation of pseudospin-1 EM wave packets travelling in the superlattice. (a) Electric field amplitude distribution of a Gaussian wave packet at $t = 0$ with the reduced center frequency of $\overline{\delta\omega_c} = 0.06\pi v_g/L$ and a half width of $r_0 = 30d$. (b) and (c) Electric field amplitude distributions at $t = 1200d/v_g$ in a single PC (left panel) and in the PC superlattice (right panel) with the initial propagation direction of the wave packet in an angle $\theta = 0^0$ (b) and $45^0$ (c), respectively, with respect to the $x$-axis. The right-hand panels show collimation behavior in the presence of the superlattice.

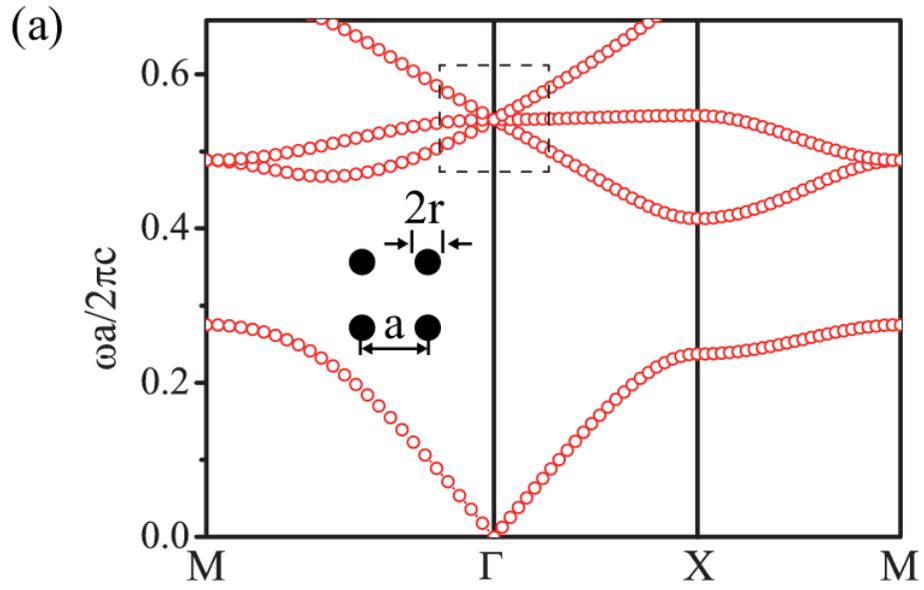

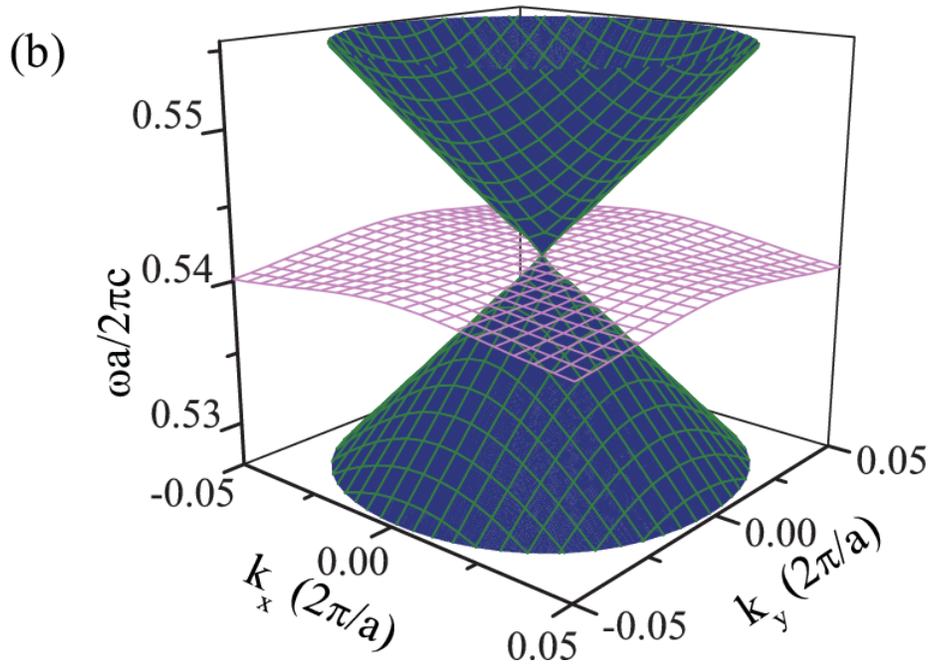

**Fig. 1**

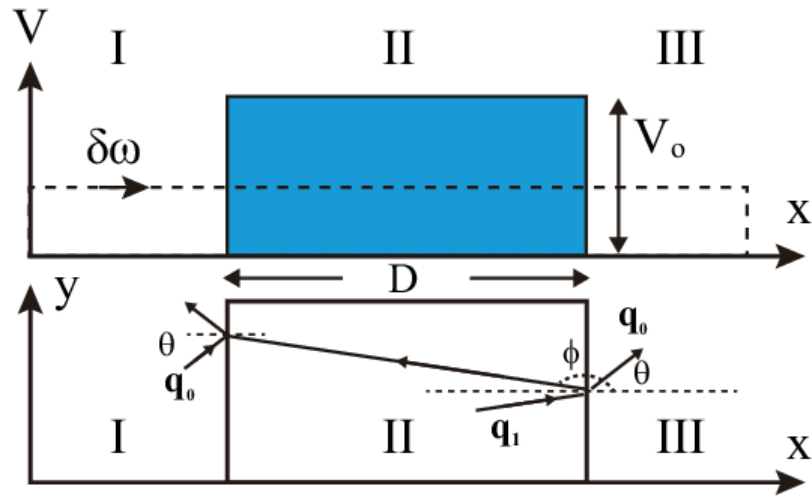

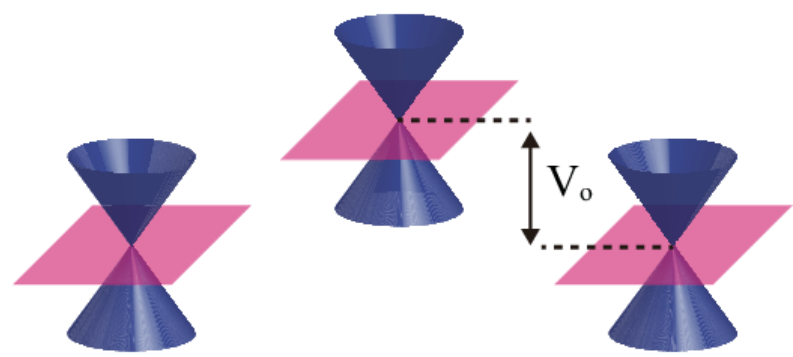

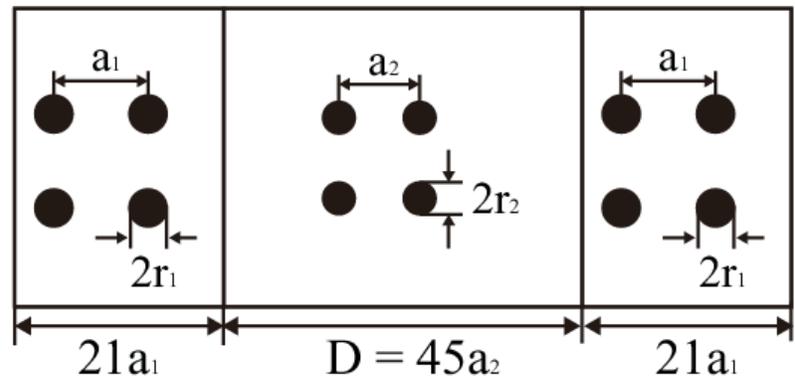

**Fig. 2**

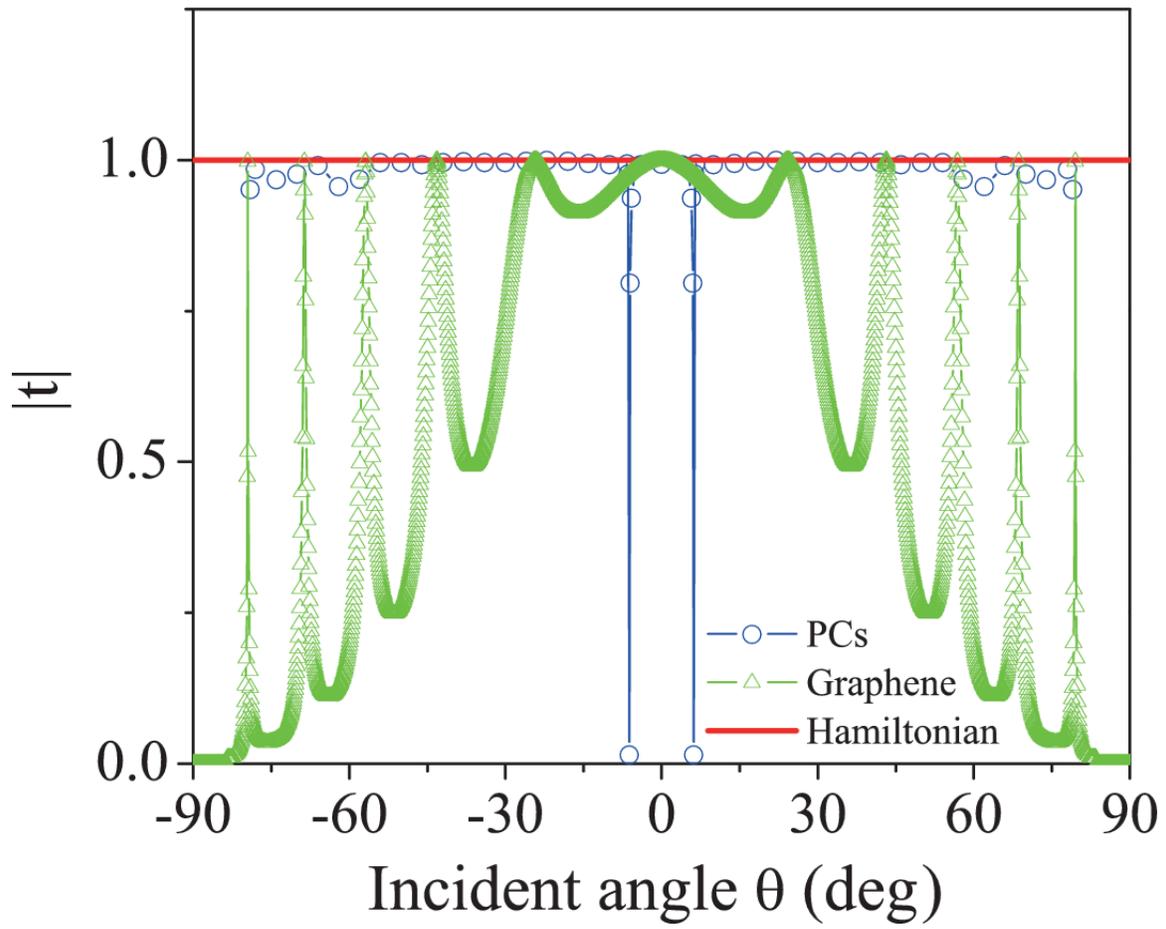

**Fig. 3**

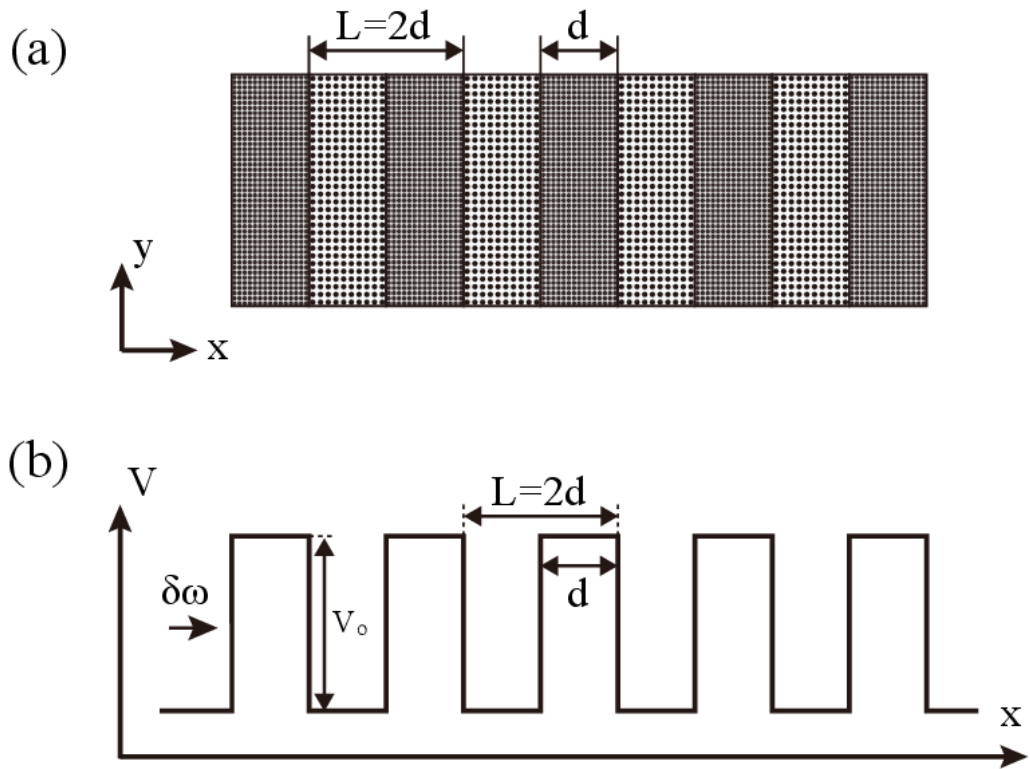

**Fig. 4**

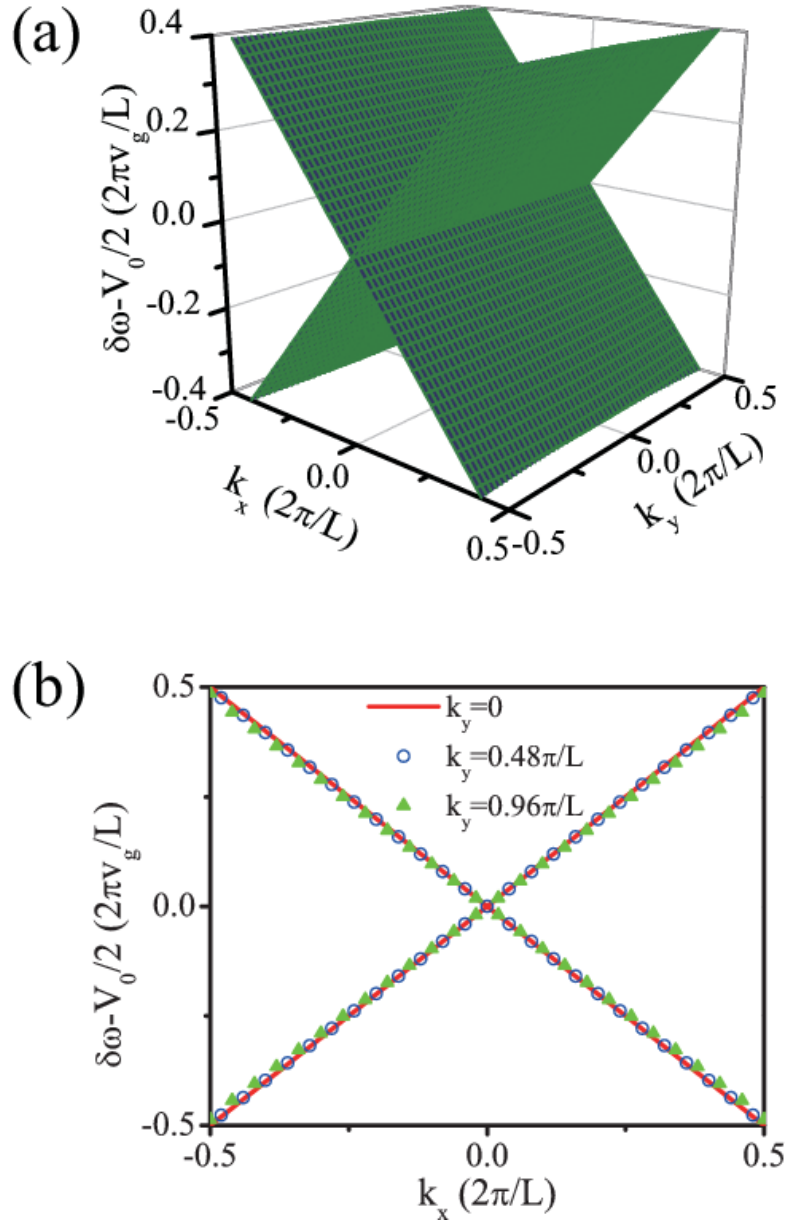

**Fig. 5**

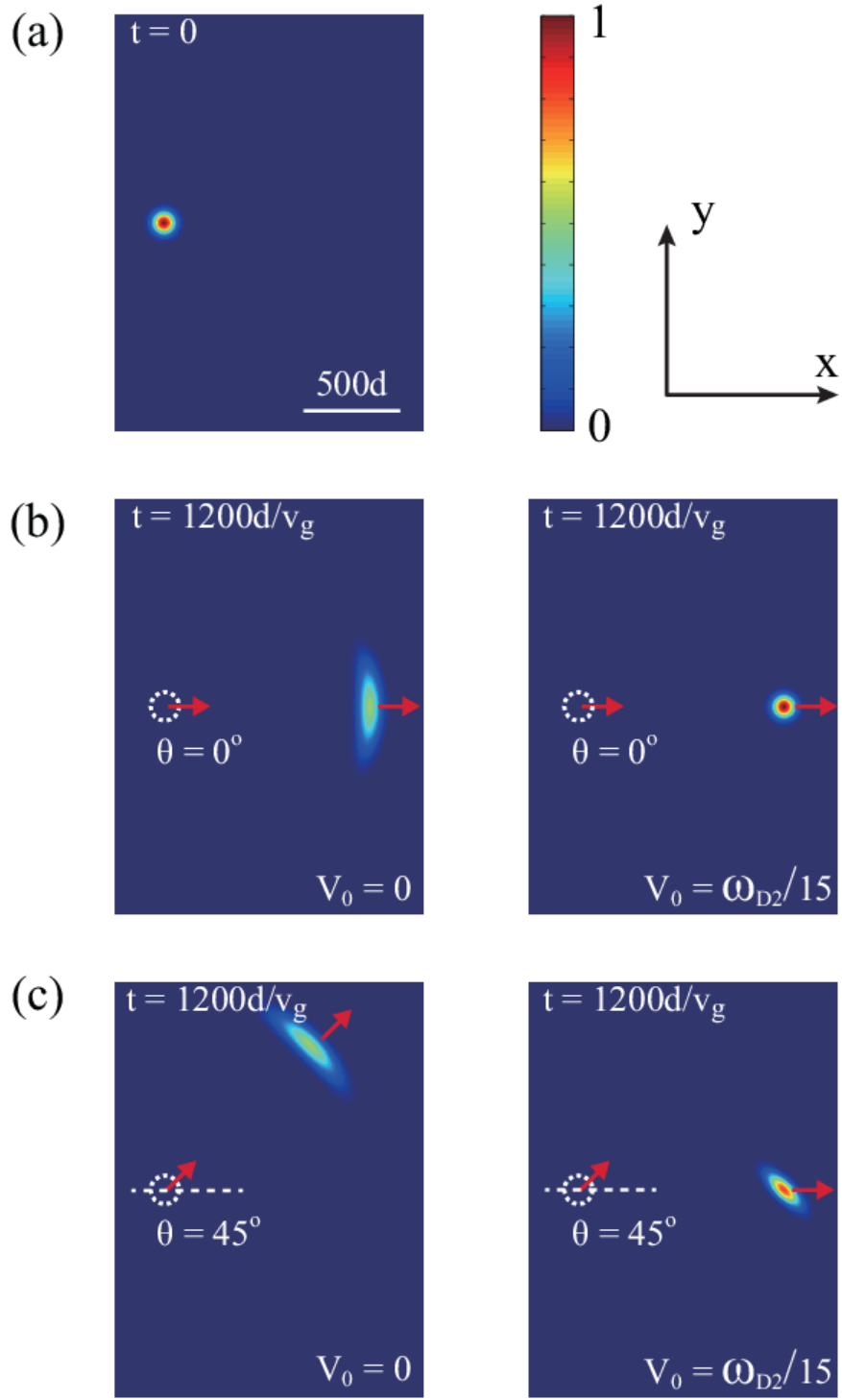

**Fig. 6**